\documentclass[aps,prd,twocolumn]{revtex4}

\begin{document}

\title{\bf Color-flavor locked strange matter}
\author{G. Lugones and J. E. Horvath\\
\it Instituto de Astronomia, Geof\'{\i}sica e Ci\^encias Atmosf\'ericas/Universidade de S\~ao Paulo\\
Rua do Mat\~ao 1226,  (05508-900) S\~ao Paulo SP, Brazil\\
Email: glugones@astro.iag.usp.br ; foton@astro.iag.usp.br}

\vskip5mm
\begin{abstract}
We analyze how the CFL states in dense matter work in the
direction of enhancing the parameter space for absolutely stable
phases (strange matter). We find that the "CFL strange matter"
phase can be the true ground state of hadronic matter for a much
wider range of the parameters of the model (the gap of the QCD
Cooper pairs $\Delta$, the strange quark mass $m_s$ and the Bag
Constant $B$) than the state without any pairing, and derive a
full equation of state and an accurate analytic approximation to
the lowest order in $\Delta$ and $m_{s}$ which may be directly
used for applications. The effects of pairing on the equation of
state are found to be small (as previously expected) but not
negligible and may be relevant for astrophysics.
\end{abstract}

\maketitle

%---------------------------------------------------------------%
\section{Introduction}
%---------------------------------------------------------------%

A great deal of activity lasting more than two decades was generated by the
hypothetic stability of strange quark matter (SQM) put forward in
Witten's seminal paper
\cite{Witten} and a few important precursors \cite{Bodmer}. These works
actually questioned the nature of the true ground state of hadronic matter and showed
within simple models that the hypothesis of a stable form of cold catalyzed
plasma was tenable. Following these works
a compreenhensive discussion of strange matter by Farhi and Jaffe \cite{farhijaffe}
in the framework of MIT Bag model of confinement \cite{MIT} presented the
so-called "windows of stability", or regions in the plane $m_{s}-B$ inside
which the stability of SQM can be realized. Other models of confinement have
also shown a fairly large range of conditions for SQM to be absolutely bound
\cite{massdep,NJL}; although it has always been clear that the
availability of a $\sim 1 \, \% $ binding energy difference for SQM to be bound
is ultimately an experimental matter.

Nevertheless, and while sophisticated experiments push the search
of SQM in laboratory and astrophysical environments beyond their
present limits, important theoretical developments have taken
place. The main one is probably the revival of interest in pairing
interactions in dense matter, a subject already addressed in the
early '80s \cite{BL} which came back a few years ago and prompted
new calculations of the pairing energy and related physics. It is
now generally agreed \cite{ARW,rap,rajagopal2} that (at least for
asymptotic densities) the color-flavor locked (CFL) state is
likely to be the ground state, even if the quark masses are
unequal \cite{uneq}. Moreover, equal number of flavors is enforced
by the symmetry, and electrons are absent since the mixture is
automatically neutral \cite{rajagopal}.

Given these important modifications in the character of the ground state
indicated by theoretical improvements, we revisit the problem of SQM in
the light of CFL state to address whether there is still room for the
Bodmer-Witten-Terazawa conjecture in Section III. Independently of stability
considerations, the equation of state for CFL matter is studied in next Section
and differences with respect to unpaired matter quantified.

%-------------------------------------------------%
\section{Thermodynamics of the CFL phase}
%-------------------------------------------------%

To order $\Delta^2$, the  thermodynamical potential
$\Omega_{CFL}$ can be found quite simply
\cite{alfordcfl}. One begins with  $\Omega_{free}$ of a fictional state of unpaired quark
matter in which all quarks which are "going to pair" have a common Fermi momentum
$\nu$, with $\nu$ chosen to minimize $\Omega_{free}$ of this fictional unpaired state.
The binding energy of the diquark condensate is included by subtracting the condensation
term $3 \Delta^2 \mu^2 / \pi^2$. Given that the mixture does not show automatic
confinement, it may be introduced at this point by means of the
phenomenological vacuum energy density or bag constant $B$. The advantages and
inconveniences of this particular implementation of confinement forces have been
discussed many times and will not be repeated here. The expression for $\Omega_{CFL}$
in this model is then \cite{alfordcfl}

\begin{eqnarray}
\nonumber \Omega_{CFL}& =& \Omega_{free} - \frac{3}{\pi^2} \Delta^2 \mu^2 + B =  \\
\nonumber&=& \frac{6}{\pi^2}  \int_{0}^{\nu} [p - \mu] p^2 dp +
\frac{3}{\pi^2}  \int_{0}^{\nu} [ (p^2 + m_s^2)^{1/2} - \mu] p^2 dp \\
& & - \frac{3}{\pi^2} \Delta^2 \mu^2 + B,\\
\nonumber &=& \sum_{i=u,d,s} \frac{1}{4 \pi^2} \bigg[ \mu_i \nu (\mu_i^2 - \frac{5}{2} m_i^2) +
\frac{3}{2} m_i^4  \log\bigg( \frac{\mu_i + \nu}{m_i} \bigg)   \bigg] \\
& & - \frac{3}{\pi^2} \Delta^2 \mu^2 + B,
\end{eqnarray}

\noindent where  $3 \mu  =  \mu_u + \mu_d + \mu_s$,
and the common Fermi momentum $ \nu = ( \mu_i^2 - m_i^2)^{1/2}$ is given by

\begin{equation}
\nu = 2 \mu - \bigg( \mu^2 + \frac{m_s^2}{3} \bigg)^{1/2} \label{nu}.
\end{equation}

\noindent The pressure, baryon number density $n_{B}$ and particle number densities are
easily derived and read

\begin{equation}
P =  - \Omega_{CFL}
\label{P}
\end{equation}

\begin{equation}
n_B  = n_u  = n_d  = n_s   =  \frac{( \nu^3  +  2 \Delta^2 \mu ) }{\pi^2}
\label{nb}
\end{equation}

\noindent Since we work at zero temperature, the energy density is given by

\begin{equation}
\varepsilon =  \sum_i \mu_i  n_i   + \Omega_{CFL} = 3 \mu  n_B  - P .
\label{E}
\end{equation}

We emphasize that, to this order, the exact nature of the interaction which generates $\Delta$
does not matter. $\Omega_{CFL}$ is given by this prescription regardless of whether
the pairing is due to a point-like four-fermi interaction,
as in NJL models, or due to the exchange of a gluon,
as in QCD at asymptotically high  energies \cite{rajagopal2}.
Of course, the strength and form of the interaction determine the value of $\Delta$, and also
its dependence with the density. Lacking of an accurate calculation for $\Delta$, which may
be as high as $\sim 100 \, MeV$, we shall keep it as a free constant parameter.

In the general case for unpaired $uds$ matter the equation of state can be derived from the
chemical potentials of Ref.\cite{farhijaffe}. As is well known, in the limit $m_{s} \rightarrow 0$
not only the particle densities become equal but also the equation of state takes the simple
form $\varepsilon = 3 P + 4 B$. Pairing introduces the $\Delta^{2}$ term in Eq. (1), thus the
equation of state picks an additional term $\varepsilon = 3 P + 4 B - (6 \Delta^{2} \mu^{2})/\pi^{2}$.
The situation is much more complicated when $m_{s} \neq 0$ because the equation of state must
be calculated numerically. However, since the mass is not large when compared to the natural
scale introduced by the chemical potential, it is generally sufficient to keep $\Omega_{free}$ to
order $m_s^4$ \cite{alfordcfl}

\begin{eqnarray}
\nonumber \Omega_{CFL} =  \frac{- 3 \mu^4}{4 \pi^2 }
   +  \frac{3 m_s^2   \mu^2}{4 \pi^2 }
        - \frac{1 -12 \log(m_s /2 \mu)}{32 \pi^2}  m_s^4  \\
- \frac{3}{\pi^2} \Delta^2 \mu^2 + B.
\end{eqnarray}

We have checked that the errors are small enough even to work to the order $m_{s}^{2}$. The
main advantage of the lowest approximation is to keep the equation of state very simple, yet
useful for most calculations, and also to make clear the effect of each parameter of the model.
To this order we have

\begin{equation}
P =  \frac{3 \mu^4}{4 \pi^2 } - \frac{3 m_s^2 \mu^2}{4 \pi^2 }
+ \frac{3}{\pi^2} \Delta^2 \mu^2 - B
\end{equation}

\begin{equation}
\varepsilon =  \frac{9 \mu^4}{4 \pi^2 } - \frac{3 m_s^2 \mu^2}{4 \pi^2 }
+ \frac{3}{\pi^2} \Delta^2 \mu^2 + B
\label{eaprox}
\end{equation}

\begin{equation}
n_B =  \frac{\mu^3}{\pi^2 } - \frac{m_s^2 \mu}{2 \pi^2 }
+ \frac{2}{\pi^2} \Delta^2 \mu
\end{equation}

\begin{equation}
\nu  =  \mu - \frac{m_s^2}{6 \mu}.
\end{equation}

Since  $m_s \sim \, 150 \, MeV$ and  $\mu$ is greater than $\sim
\,  300 MeV$  this approximation is quite accurate, especially at
high densities, as is apparent from Fig. 1.

It is also desirable to have an  expression of $P$ as an explicit
function of $\varepsilon$. From Eqs. (\ref{eaprox}) and (\ref{P})
we obtain

\begin{equation}
\varepsilon =   3P + 4B
- \frac{6 \Delta^2 \mu^2}{\pi^2} +  \frac{3 m_s^2 \mu^2}{2 \pi^2 }
\label{eos}
\end{equation}

\noindent where $\mu^2$ is given by

\begin{equation}
\mu^2 = - \alpha + \bigg( \alpha^2 + \frac{4}{9} \pi^2 (\varepsilon - B) \bigg)^{1/2}
\label{mu2}
\end{equation}

\noindent and

\begin{equation}
\alpha = - \frac{m_s^2}{6} +  \frac{2 \Delta^2}{3}.
\end{equation}

Equation (\ref{eos}) resembles the EOS for strange quark matter
with massless quarks with the addition of  the last two terms. The
term proportional to $\Delta^2$ tends to stiffen the EOS compared
to the SQM case since induces a higher pressure for  a given
energy density. The term with  $m_s^2$ has the opposite effect,
although it is not as large. The CFL state may be preferred to SQM
in spite of the finite $m_{s}$ value because of the importance of
the $\Delta$ term. The effect of color flavor locking in the
equation of state is not negligible although it is not extreme
either. Given that  $\Delta \, \sim 100$  MeV and that a typical
$\mu$ is $\geq 300$ MeV the effect of CFL in the EOS  may be
important, specially at low densities. We show in Figure 1  the
EOS in the different approximations. From the expressions above,
it is readily noticed that, provided $\Delta$ is higher than
$m_{s}/2$, the EOS is stiffer than the SQM, that is, produces more
pressure for a given energy density. Since the actual value of
$\Delta$ is not well known, we expect either a stiffer or a softer
EOS (for a given $B$). It should be kept in mind that there are
other caveats, for example, the likely dependence of $\Delta$ on
the density, which may cause a cross from stiffer to softer EOS
depending on the parameters.

%------------------------------------------------------
\section{Stability of the CFL phase}
%------------------------------------------------------

For a given EOS the energy per baryon of the deconfined phase (at
$P=0$ and $T=0$) must be lower than $939 \, MeV$ (the neutron
mass) if matter is to be absolutely stable. The other condition
that must be considered comes from the empirically known stability
of normal nuclear matter against deconfinement at zero pressure
\cite{farhijaffe}. In other words the energy per baryon of
deconfined matter (a pure gas of quarks $u$ and $d$) at zero
pressure and temperature must be higher than the neutron mass
value. In the framework of a MIT-based EOS it has been shown that
the latter condition imposes that the MIT Bag Constant must be
greater than $57 \, MeV \, fm^{-3}$ \cite{farhijaffe}.

From Eq. (\ref{E}) we can write the absolute stability condition as

\begin{equation}
\frac{\varepsilon}{ n_B} \bigg|_{P = 0} =  3 \mu   \leq  m_n = 939 MeV .
\end{equation}

This  simple result is a direct consequence of the existence of a common Fermi momentum
for the three flavors and is valid at $T = 0$ without any approximation.
Since this must hold at the zero pressure point, then, from Eq. (\ref{P}) we have

\begin{equation}
B =  - \Omega_{free}(m_s,\mu_0) + \frac{3}{\pi^2} \Delta^2 \mu_0^2
. \label{window}
\end{equation}

\noindent with  $\mu_0 = {313} \, MeV$.

The last equation defines a curve in the $m_s - B$ plane on which
the energy per baryon is exactly $\varepsilon / n_B = m_n$ for a given $\Delta$.
To order $m_s^2$ we can obtain a very simple parabolic
expression for Eq. (\ref{window}):

\begin{equation}
B =  - \frac{m_s^2 m_n^2}{12 \pi^2} + \frac{\Delta^2 m_n^2}{3 \pi^2}
+  \frac{m_n^4}{108 \pi^2}
\label{windowanalytic}
\end{equation}

Since this analytic expression is calculated to order $m_s^2$, it deviates
from Eq.(\ref{window}) when $m_{s} \sim \mu$, in practice the approximation holds for
masses up to about $150 \, MeV$, expected to be quite realistic.

We display in Fig. 2 the stability window for the CFL phase (i.e.
the region in the $m_s$ versus B plane where $E/n_B$ is lower than
$939 \, MeV$ at zero pressure. Eq. (\ref{window}) gives the right
side boundary of the window while the left side boundary is given
by the minimum value $B = 57 \, MeV$. As it stands, the window is
greatly enlarged for increasing values of $\Delta$. This is to be
compared, for example, with Fig. 1 of Ref. \cite{farhijaffe} in
which no pairing was included. The $\Delta$ term actually produces
this effect of enlargement of the parameter space.

%------------------------------------------------------
\section{Discussion}
%------------------------------------------------------

The CFL phase at zero temperature has been modelled as an
electrically neutral and colorless fluid, in which quarks are
paired in such a way that all the flavors have the same Fermi
momentum and hence the same number density, as long as $m_s$ is
not too large \cite{rajagopal}. The CFL phase is strongly favored
over a pure mixture of quarks $u$ and $d$ and pure neutron matter
for a wide range of parameters of the theory (namely $B$, $m_s$
and $\Delta$). Although some energy must be paid in order to
maintain the same Fermi momentum for all three flavors, more
energy is gained by opening the strange quark channel and from the
energy gap of the pairing.

The CFL phase treated here as a gas of Cooper pairs following
Ref.\cite{alfordcfl} shows a (qualitatively and quantitatively)
different behavior to that developed for quark-diquark matter in
\cite{vigilantes}, where diquarks are treated as bosons much in
the same way as Refs.\cite{Donoghue,Kastor,HP}. In the latter
case, the effect of Bose condensation is much more important than
the energy gap of the pairing itself, while in the present case
the gap energy is essential to widen the stability window. The gap
effect does not dominate the energetics, being of the order
$(\Delta/\mu)^{2}$ (a few percent), but it may be large enough to
allow a "CFL strange matter" for the same parameters that would
otherwise produce unbound strange matter without pairing. Similar
conclusions have been recently presented by Madsen \cite{Madsen}
in a study focused on CFL strangelets (not addressed here). We
believe that the explicit analytic expressions derived in section
II may be useful to study strange stars and related problems,
while Fig. 2 quantifies the expected enlargement of the stability
windows in a convenient manner for comparison with "ordinary" SQM
\cite{farhijaffe}.

Even if the EOS is very simple and the confinement has been
introduced by brute force, it is remarkable that the strange
matter hypothesis may be boosted by pairing interactions, and
clearly more detailed studies are desirable. The dynamics of the
transition itself is also a matter of interest. While it is likely
that a 2SC phase may be bypassed in favor of a CFL state
\cite{trol}, the original flavor content of the hadronic phase is
generally not the one needed by the CFL flavor symmetry.
Therefore, it is still reasonable to assume that the transition
dynamics is dominated by the rate of strangeness production needed
to achieve the CFL flavor symmetry. The energy liberated in the
transition from the hadronic to CFL state could be much higher
than that liberated in the process of unpaired SQM formation and
could lead to very energetic explosive phenomena
\cite{Comb,chin,hsu}. It is also worth to remark that stable
diquark states have been suggested some time ago although within a
different (naiver) model \cite{SDM}.

%----------------------------------------------------------------
\section{Acknowledgements} G. Lugones  acknowledges the Instituto de Astronomia,
Geof\'\i sica e Ci\^encias Atmosf\'ericas de S\~ao Paulo and the financial
support received from
the Funda\c c\~ao de Amparo \`a Pesquisa do Estado de
S\~ao Paulo. J.E. Horvath wishes to acknowledge
the CNPq Agency (Brazil) for partial financial support.

%---------------------------------------------------------------%

\begin{figure}
\caption{The EOS for CFL SQM and for SQM without color flavor
locking. We have chosen B=75 MeV fm$^{-3}$ and $m_S$ = 150 MeV for
all the curves, which are shown for two different values of the
gap $\Delta$ as indicated in the figure. The solid line
corresponds to SQM (no CFL); the dashed lines are the CFL
calculated to all orders in $m_S$ and the dotted lines are the
approximate EOS to the order $m_{S}^{2}$, which results quite
accurate. Note the change of stiffness according to the value of
$\Delta$, as discussed in the text.}
\end{figure}

\begin{figure}
\caption{The windows of stability for CFL strange matter. The
symmetric CFL state is absolutely bound if the strange quark mass
$m_{s}$ and the vacuum energy density $B$ lie inside the bounded
region. Each window has been calculated for a given value of the
gap $\Delta$ as indicated by the label, to be compared with the
SQM results of \cite{farhijaffe}. The solid lines are calculated
to all orders in $m_{S}$, while the dashed lines are the
approximate regions to order $m_{S}^{2}$ as given by
Eq.(\ref{windowanalytic}). As expected, the approximation is worse
for increasing values of $m_{S}$. The vertical solid line is the
limit imposed by requiring instability of two-flavor quark matter.
The large increase of the stable region is the main feature of
interest.}
\end{figure}

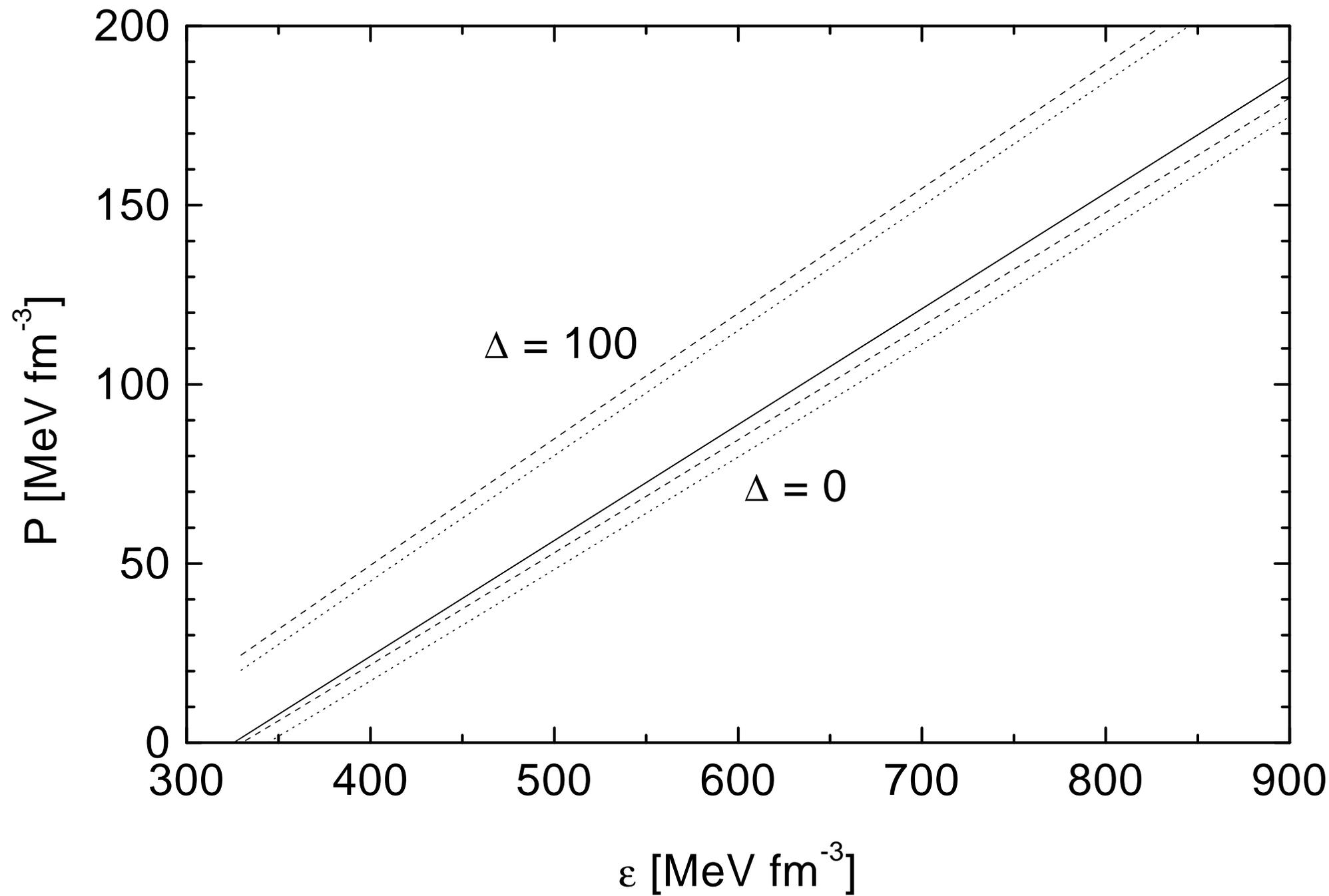
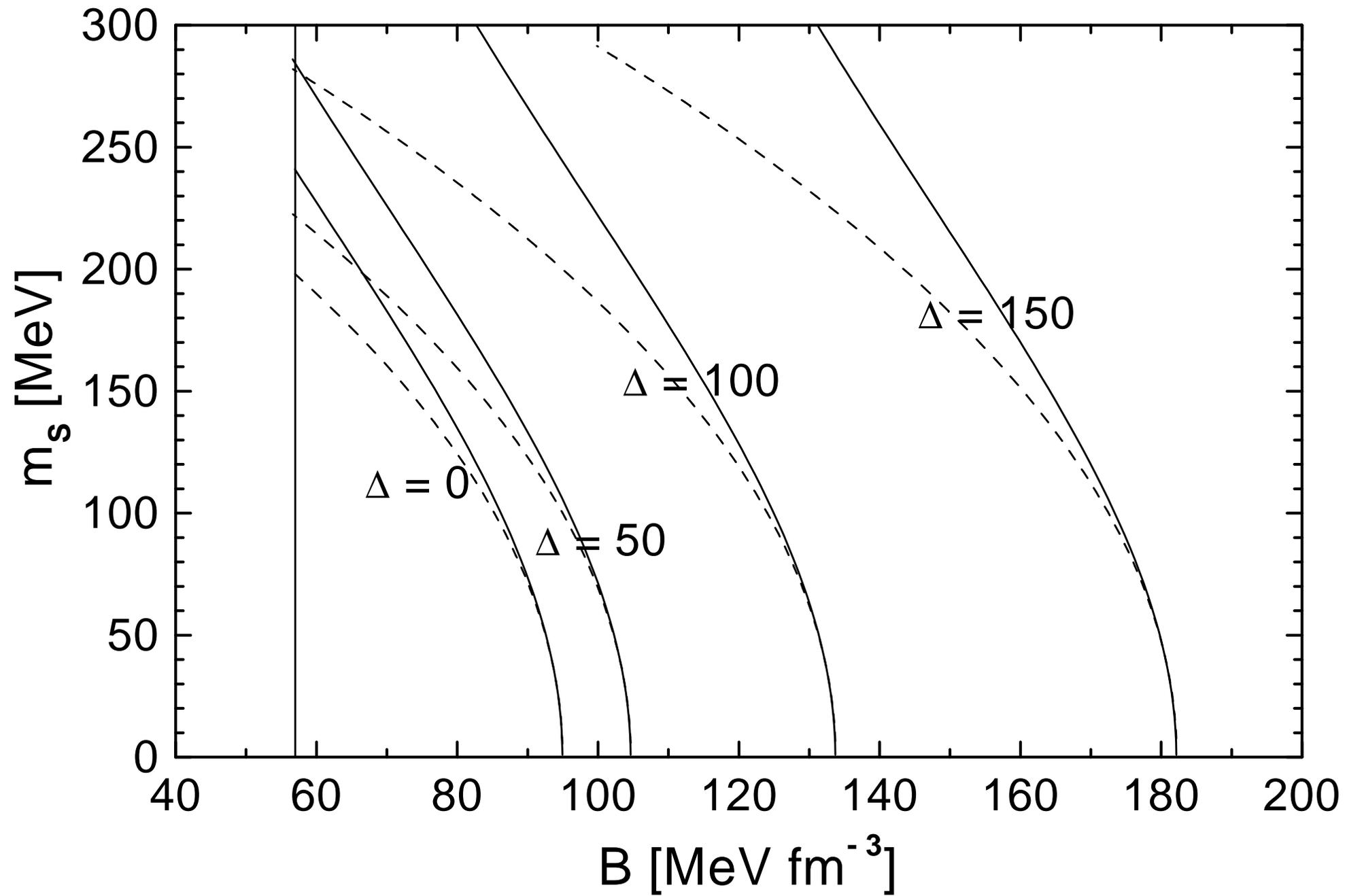
\end{document}